\documentclass[onecolumn,pra]{revtex4}
\usepackage{graphicx, amssymb, amsmath}

\begin{document}
\title{Preparation of arbitrary qutrit state based on biphotons}

\author{Yu.I.Bogdanov}

\affiliation{FTIAN, Quantum computer physics laboratory, 117218,
Moscow, Russia;}
\author{M.V.Chekhova, S.P.Kulik, G.A.Maslennikov,
} \affiliation{Department of Physics, Moscow M.V. Lomonosov State
University, 119992 Moscow, Russia.}
\email{postmast@qopt.phys.msu.su}
\author{C.H.Oh, M.K.Tey}
\affiliation{Department of Physics, Faculty of Science, National
University of Singapore, 117542 Singapore.}
\date{\today}


\begin{abstract}
The novel experimental realization of three-state optical quantum
systems is presented. We use the polarization state of
biphotons,propagating in single frequency- and spatial mode, to
generate an arbitrary qutrits. In particular the specific sequence
of states that are used in the extended version of BB84 quantum
key distribution protocol was generated and tested. We
experimentally verify the orthogonality of the 12 basic states and
demonstrate the ability of switching between them. The tomography
procedure is applied to reconstruct the density matrices of
generated states.
\end{abstract}
\maketitle
\section{INTRODUCTION}
\subsection{Three-state systems in Quantum Information}
  The art of quantum state engineering, i.e., the ability
to generate, transmit and measure quantum systems is of great
importance in the emerging field of quantum information
technology. A vast majority of protocols relying on the properties
of two-level quantum systems (qubits) were introduced and
experimentally realized. But naturally, there arose a question of
an extension of dimensionality of systems used as information
carriers and the new features that this extension can offer. The
simplest extension provokes the usage of three-state quantum
systems (qutrits). Recently new quantum key distribution (QKD)
protocols were proposed that dealt specifically with qutrits
\cite{peres:00,kasz:03} and the eavesdropping analysis showed that
this systems were more robust against specific classes of
eavesdropping attacks \cite{bruss:02,durt:03} . The other
advantage of using multilevel systems is their possible
implementation in the fundamental tests of quantum mechanics
\cite{collins:02} , giving more divergence from classical theory.
The usage of multilevel systems also provides a possibility to
introduce very specific protocols, which cannot be implemented
with the help of qubits such as Quantum Bit Commitment, for
example \cite{lang:04} . Recent experiments on realization of
qutrits rely on several issues. In one case, the interferometric
procedure is used, where entangled qutrits are generated by
sending an entangled photon pair through a multi-armed
interferometer \cite{rob:01} . The number of arms defines the
dimensionality of the system. Other techniques rely on the
properties of orbital angular momentum of single photons
\cite{lang:04,zei:01,zei:03} and on postselection of qutrits from
four-photon states \cite{antia:01} . Unfortunately all mentioned
techniques provide only a partial control over a qutrit state. For
example in a method, mentioned in \cite{lang:04,zei:01,zei:03} a
specific hologram should be made for given qutrit state. The real
parts of the amplitudes of a qutrit, generated in \cite{rob:01}
are fixed by a characteristics of a fiber tritter, making it hard
to switch between the states. Besides, in this method no
tomographic control over generated state had been yet performed.

\subsection{Biphoton properties}
In this paper we report the experimental realization of arbitrary
qutrit states that exploits the polarization state of single-mode
biphoton field. This field consists of pairs of correlated
photons, is most easily obtained with the help of spontaneous
parametric down-conversion (SPDC). By saying "single-mode" we mean
that twin photons forming a biphoton have equal frequencies and
propagate along the same direction. A pure polarization state of
such field can be written as the following superposition of three
basic states.
\begin{equation}
\begin{array} {cc}
|\Psi\rangle=c_{1}|2,0\rangle+c_{2}|1,1\rangle+c_{3}|0,2\rangle,
\label{eq:state}
\end{array}
\end{equation}
where $c_i=|c_i|e^{i\phi_i}$ are the complex probability
amplitudes. Bases states are the Fock states with certain number
of photons in two orthogonal polarisation modes. It is reasonable
to use vertical and horizontal polarisation modes for basis state
definition. So by state $|2,0\rangle$ we mean that two photons
have horizontal polarization, $|0,2\rangle$ is the basis state
were both of the photons are polarized vertically and
$|1,1\rangle$ is the basis state with one horizontally polarized
photon and one vertically polarized photon. Such a states can be
generated in SPDC processes, first two via type I phase-matching
SPDC and the last one via type II phase-matching SPDC.

 There exists an alternative
representation of state $|\Psi\rangle$ that maps the state onto
the surface of Poincare sphere \cite{masha:01} \vspace{0.5cm}
\begin{equation}\footnotesize
|\Psi\rangle=\frac{a^\dagger(\theta,\phi)a^\dagger(\theta^\prime,\phi^\prime)|vac\rangle}{\parallel
a^\dagger(\theta,\phi)a^\dagger(\theta^\prime,\phi^\prime)|vac\rangle
\parallel} \label{eq:state2}
\end{equation}
where $a^\dagger(\theta,\phi)$ and
$a^\dagger(\theta^\prime,\phi^\prime)$ are the creation operators
of a photon in the certain polarization mode
$a^\dagger(\theta,\phi)=cos(\theta/2)a_x^\dagger+exp(i\phi)sin(\theta/2)a_y^\dagger$,
$a_x^\dagger,a_y^\dagger$ are photon creation operators in
correspondingly horizontal and vertical polarization modes,
$\theta,\theta^\prime\in[0,\pi]$, $\phi,\phi^\prime\in[0,2\pi]$
are polar and azimuthal angles that define the position of each
photon on a surface of a sphere. The values of the angles can be
calculated using amplitudes and phases of $c_{i}$.

\begin{figure}[!ht]
\begin{center}
\begin{tabular}{c}
\includegraphics[
height=4cm ]{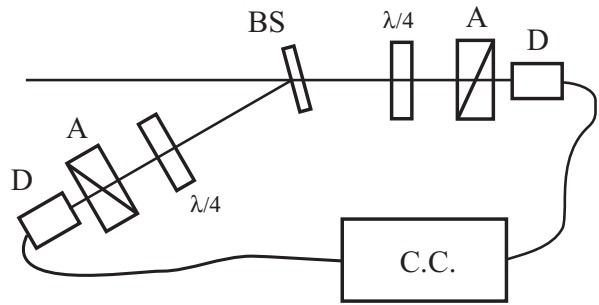}
\end{tabular}
\end{center}
\caption{\label{fig:orthoc} Brown-Twiss interferometer with
polarization selection.}
\end{figure}

The operational orthogonality criterion for the polarization
states of single-mode biphotons was proposed in \cite{masha:02}
and experimentally verified in \cite{ortexp} .
 Let us consider setup shown on the Fig.~\ref{fig:orthoc} in
 order to formulate the criterion.
Presented setup is a
 Brown-Twiss interferometer with a quarter-wavelength plate and an analyzer in each arm.
 Both the plate and the analyzer in one arm could be rotated independently by an arbitrary
 angle.  According to these angles there is a certain polarization of photons that corresponds
 to their lossless propagation through the arm. Let us assume that
 these polarization photon states are $a^\dagger(\theta,\phi)|vac\rangle$ for
 transmitted arm and $a^\dagger(\theta^\prime,\phi^\prime)|vac\rangle$ for reflected arm.
 In this case we would say that setup is "tuned" to detect the
 biphotons in polarization state $|\Psi\rangle$
 (~\ref{eq:state2}).

 If we have the biphotons in state $|\Psi\rangle$ in the input of the
 setup the coincidence count rate would be non-zero. The most
 interesting question is what happen with coincidence count rate $R_c$ when the biphotons in the
 input of the setup that is tuned to state $|\Psi\rangle$ are in
 any other state $|\Psi^\prime\rangle$.
The answer was obtained in \cite{masha:02} where was shown that
$R_c \sim | \langle \Psi|\Psi^\prime \rangle |^2$. So this scheme
is a realization of the projector of an input state on the fixed
one. In particular if we have input state orthogonal to the one
setup tuned there would not be any coincidences in the scheme.
Thus the coincidence rate is zero if and only if the input state
is orthogonal to the one setup tuned. The significance of this
criterion would be shown at the next section. \par The goal of our
work was to demonstrate the ability to prepare any given
polarization state $|\Psi\rangle$ and as a straightforward and
practical example of given states, we chose the specific sequence
that was presented in \cite{peres:00} . This sequence of 12 states
can be used in an extended version of BB84 QKD protocol for
qutrits. As it was shown one can construct 4 mutually unbiased
bases with 3 orthogonal states in each from any three orthogonal
states using discrete Fourier transformation. Using
$\alpha=|2,0\rangle$, $\beta=|1,1\rangle$, $\gamma=|0,2\rangle$ as
the first orthogonal basis we observed all 12 states shown in the
table I.
\begin{table}[!ht]
\caption{12 states used in qutrit QKD protocol.}
\begin{center}
\begin{tabular}{|l|l|l|l|l|l|l|}
\hline
State&$|c_1|$&$|c_2|$&$|c_3|$&$\phi_1$&$\phi_2$&$\phi_3$\\\hline
$|\alpha\rangle$&$1$&$0$&$0$&$0$&$0$&$0$\\
\hline
$|\beta\rangle$&$0$&$1$&$0$&$0$&$0$&$0$\\
\hline
$|\gamma\rangle$&$0$&$0$&$1$&$0$&$0$&$0$\\
\hline
$|\alpha^\prime\rangle$&$\frac{1}{\sqrt{3}}$&$\frac{1}{\sqrt{3}}$&$\frac{1}{\sqrt{3}}$&$0$&$0$&$0$\\
\hline
$|\beta^\prime\rangle$&$\frac{1}{\sqrt{3}}$&$\frac{1}{\sqrt{3}}$&$\frac{1}{\sqrt{3}}$&$0$&$120^\circ$&$-120^\circ$\\
\hline
$|\gamma^\prime\rangle$&$\frac{1}{\sqrt{3}}$&$\frac{1}{\sqrt{3}}$&$\frac{1}{\sqrt{3}}$&$0$&$-120^\circ$&$120^\circ$\\
\hline
$|\alpha^{\prime\prime}\rangle$&$\frac{1}{\sqrt{3}}$&$\frac{1}{\sqrt{3}}$&$\frac{1}{\sqrt{3}}$&$120^\circ$&$0$&$0$\\
\hline
$|\beta^{\prime\prime}\rangle$&$\frac{1}{\sqrt{3}}$&$\frac{1}{\sqrt{3}}$&$\frac{1}{\sqrt{3}}$&$0$&$120^\circ$&$0$\\
\hline
$|\gamma^{\prime\prime}\rangle$&$\frac{1}{\sqrt{3}}$&$\frac{1}{\sqrt{3}}$&$\frac{1}{\sqrt{3}}$&$0$&$0$&$120^\circ$\\
\hline
$|\alpha^{\prime\prime\prime}\rangle$&$\frac{1}{\sqrt{3}}$&$\frac{1}{\sqrt{3}}$&$\frac{1}{\sqrt{3}}$&$-120^\circ$&$0$&$0$\\
\hline
$|\beta^{\prime\prime\prime}\rangle$&$\frac{1}{\sqrt{3}}$&$\frac{1}{\sqrt{3}}$&$\frac{1}{\sqrt{3}}$&$0$&$-120^\circ$&$0$\\
\hline
$|\gamma^{\prime\prime\prime}\rangle$&$\frac{1}{\sqrt{3}}$&$\frac{1}{\sqrt{3}}$&$\frac{1}{\sqrt{3}}$&$0$&$0$&$-120^\circ$\\
\hline
\end{tabular}
\end{center}
\end{table}
where $\alpha$, $\beta$ and $\gamma$ with fixed number of primes
are the orthogonal states belong to the same basis and a scalar
product of any two states belong to different bases is equal to
1/3.

\section{EXPERIMENT}

\subsection{Experimental setup and method of measurement}

Our setup consists of two parts: first one performed the
preparation of the biphoton field in a desired state and second
one provide the measurement of the obtained states.

\begin{figure}[!ht]
\begin{center}
\begin{tabular}{c}
\includegraphics[
height=4cm ]{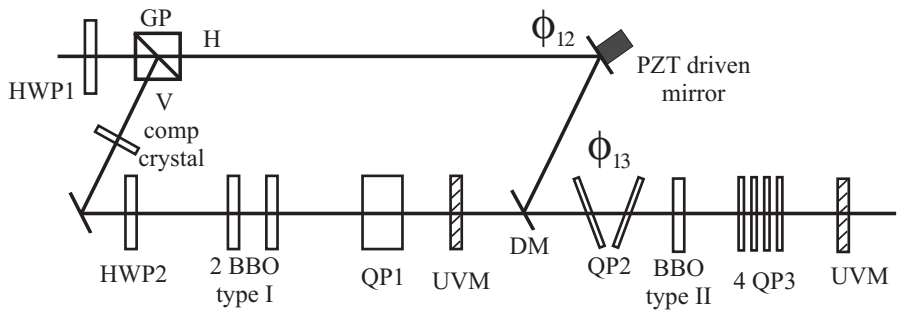}
\end{tabular}
\end{center}
\caption{\label{fig:setup1} Experimental setup (preparation
part).}
\end{figure}

The preparation part of our setup (Fig.~\ref{fig:setup1}) is built
on the base of a balanced Mach-Zehnder interferometer
    (MZI) \cite{gleb:03} .

The pump part consists of frequency doubled "Coherent Mira 900"
femtosecond laser, operated at central wavelength of 800 nm, 75
MHz repetition rate and with a pulse width of 100 fs, average pump
power was 20 mW. The Glan-Tompson prism (GP), transmitting the
horizontally polarized fraction of the UV pump and reflecting the
vertically polarized fraction, serves as an input mirror of MZI.
The reflected part, after passing the compensation BBO crystal and
a half-wave plate (HWP2), pumps two consecutive 1 mm thick type-I
BBO crystals whose optical axis are oriented perpendicularly with
respect to each other. The biphotons from these crystals pass
through a 10 mm quartz plate (QP1) that serves as a compensator,
and the pump is reflected by an UV mirror. Then the biphotons
arrive at a dichroic mirror (DM) that is designed to transmit them
and to reflect the horizontally polarized component of the pump
coming from the upper arm of MZI. A piezoelectric translator (PZT)
was used to change the phase shift of the horizontal component of
the pump with respect to the one propagating in the lower arm. The
UV beam, reflected from DM serves as a pump for 1 mm thick type-II
BBO crystal. Two 1 mm quartz plates (QP2) can be rotated along the
optical axis to introduce a phase shift between horizontally and
vertically polarized type-I biphotons, and a set of four 1 mm
thick quartz plates (QP3) serves to compensate the group velocity
delay between orthogonally polarized photons during their
propagation in type II BBO crystal.

\begin{figure}[!ht]
\begin{center}
\begin{tabular}{c}
\includegraphics[
height=3cm ]{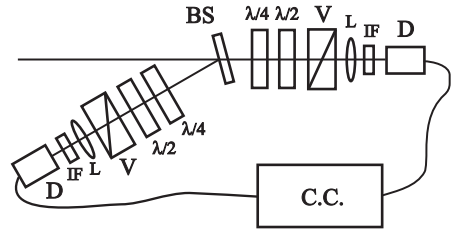}
\end{tabular}
\end{center}
\caption{\label{setup2} Experimental setup (measurement part).}
\end{figure}

The measurement setup (Fig.~\ref{setup2}) consists of a
Brown-Twiss scheme with a non-polarizing 50/50 beamsplitter; each
arm contains consecutively placed quarter- and half waveplates and
an analyzer that was set to transmit the vertical polarization.

This sequence of waveplates and analyzer is referred to as a
polarization filter as soon as it extracts a single-photon
polarization state defined by their orientation. Interference
filters of 5 nm bandwidth, centered at 800 nm and pinholes are
used for spectral and spatial modal selection of biphotons. We use
EGG-SPCM-AQR-15 single photon counting modules as our detectors
(D1 and D2). We should mention, that due to the low pump power,
the stimulated processes in our setup are negligibly small and
only pairs of photons have been generated. The measurement of the
generated states is done using the tomography protocol that was
developed for polarization qutrits \cite{leo:04,bogdan:03a} . From
experimental point of view we have to measure all fourth-order
correlation moments in order to reconstruct coherence matrix
 \begin{center}
$K_{4}=
$$
\begin{pmatrix}
A & D & E \\
D^{*} & C & F \\
E^{*} & F^{*} & B
\end{pmatrix}
$$
$\end{center}

 which contains the following moments
$
\begin{array}{ccc}
     A=\langle a^{\dagger^2}a^2\rangle,& B=\langle b^{\dagger^2}b^2\rangle,& C=\langle a^{\dagger}b^{\dagger}ab\rangle, \\
     D=\langle a^{\dagger^2}ab\rangle ,& E=\langle a^{\dagger^2}b^2\rangle,& F=\langle a^{\dagger}b^{\dagger}b^2\rangle.
\end{array}
$

In the experiment one can measure directly only the diagonal
elements of the coherency matrix. For example one measures B
moment when there are only two vertical polarizers in the arms and
no plates , adding quarter wavelength plate rotated at $45^{0}$ to
one of the arms - C moment, adding quarter wavelength plates
rotated at $45^{0}$ to each arm - A moment. In all other cases the
coincidence rate would be proportional to linear combination of
different moments. By choosing an appropriate combination
containing as few number of moments as possible was proposed the
following tomography protocol shown on the table 2.

\begin{table}[!ht]
\caption{Tomography protocol. Each line contains orientation of
the half ($\theta_{s,i}$) and quarter ($\chi_{s,i}$) wave plates
in the measurement part of the experimental setup. Last column
show the corresponding moment.}
\begin{center}
\begin{tabular}{|c|c|c|c|c|c|c|}
 \hline
 & \multicolumn{4}{c|}{Parameters of the experimental set-up}
 & Moments to be measured\\\hline
  $\nu$&$\chi_s$&$\theta_s$&$\chi_i$&$\theta_i$& \\\hline
  1. & 0 & $45^\circ$ & 0 & $-45^\circ$ & $A/4$ \\\hline
  2. & 0 & $45^\circ$ & 0 & 0 & $C/4$ \\\hline
  3. & 0 & 0 & 0 & 0 & $B/4$ \\\hline
  4. & $45^\circ$ & 0 & 0 & 0 & $1/8(B+C+2\textrm{Im}F)$ \\\hline
  5. & $45^\circ$ & $22.5^\circ$ & 0 & 0 & $1/8(B+C-2\textrm{Re}F)$  \\\hline
  6. & $45^\circ$ & $22.5^\circ$ & 0 & $-45^\circ$ & $1/8(A+C-2\textrm{Re}D)$ \\\hline
  7. & $45^\circ$ & 0 & 0 & $-45^\circ$ & $1/8(A+C+2\textrm{Im}D)$ \\\hline
  8. & $-45^\circ$ & $11.25^\circ$ & $-45^\circ$ & $11.25^\circ$ & $1/16(A+B-2\textrm{Im}E)$ \\\hline
  9. & $45^\circ$ & $22.5^\circ$ & $-45^\circ$ & $22.5^\circ$ & $1/16(A+B-2\textrm{Re}E)$  \\ \hline
\end{tabular}
\end{center}
\end{table}

As a result of the series of 9 measurements with different
orientation of the plates in the arms of Brown-Twiss
interferometer we can reconstruct the coherency matrix. Since we
deal with a three-dimensional Hilbert space density matrix of
measured state determined by 8 real numbers in general case when
the state is mixed and by 6 real numbers in case of the pure
state. In the experiment we have to provide one more measurement
in order to find the normalization. Working with  single mode
biphoton polarization states we can completely describe them by
means of the $K_{4}$. Since the coherence matrix carry full
information about the state it can be used as well as the density
matrix. But density matrix calculations are more usual, so we are
going to use them. Polarization density matrix can be found from
correlation moments by the formula \cite{dnk:97,bogdan:03a}

\begin{center}
\begin{equation}
\rho=
\begin{pmatrix}
2A & \sqrt{2}D & 2E \\
\sqrt{2}D^{*} & C & \sqrt{2}F \\
2E^{*} & \sqrt{2}F^{*} & 2B
\end{pmatrix} \label{eq:rho}
\end{equation}
\end{center}

Thus, after 9 measurements with the different orientations of the
plates in the setup we have ehough information to reconstruct the
density matrix of the polarization state. Moreover, this
configuration of the measurement setup (fig.~\ref{setup2}) allows
us to verify the orthogonality of the states that belong to the
same basis.

   \subsection{Compensation}
    In order to have the three terms in superposition (\ref{eq:state}) interfering, one must achieve their
    perfect overlap in frequency, momentum and time domains. From the experimental point of view this means that
    the biphoton wavepackets coming from the two type I crystals and from the type II crystal must be overlapped.
    The overlap in the frequency domain is achieved by the usage of 5nm bandwidth interference filters and the overlap
    in momentum is ensured by using pinholes that select one spatial mode of the biphoton field. But the overlap
    in time cannot be achieved easily when using a pulsed laser source, because it is necessary to compensate for all
    the group delays that biphoton wavepackets acquire during their propagation through the optical elements of the
    setup \cite{sergei:00}. It was found that in order to overlap type-I biphotons with type-II, the pump pulse from
    the lower arm must be delayed. In our case the value of the delay is ~50 ps.
    This was achieved by inserting an additional 2 mm BBO crystal in the lower arm.
    The overlap between the states $|2,0\rangle$ and $|0,2\rangle$ was achieved by inserting a 10 mm quartz plate
    directly after the two type I BBO crystals.  After overlapping the biphotons with these techniques, the average
    coincidence count rate that we observed was of about 1 Hz. The high visibility of interference patterns that we obtained
    was a criterion for a good compensation (see figures Fig.~\ref{STangle}, Fig.~\ref{STphase}, Fig.~\ref{PolIntf} in next section).

 \subsection{Calibration}
     In order to create a given qutrit state we needed to have independent control over four real parameters -
    two relative amplitudes and two relative phases.
    \par {\bf Real amplitudes.} In the experiment we used HWP1 to control the amplitude of the
    state $|1,1\rangle$, and HWP2 to control the relative amplitudes of the states $|2,0\rangle$ and $|0,2\rangle$.
    The calibration of these elements can be done by measuring moments related to $\rho_{11}, \rho_{22},
    \rho_{33}$, defined in eq. ~(\ref{eq:rho}) in the tomography setup \cite{leo:04} . For our concrete experiment we had equal weights of each basis state
    in the superposition, so we had to set our waveplates in such a way that these moments fulfilled the condition:
    $\rho_{11}=\rho_{22}=\rho_{33}$. The values of the diagonal density matrix components were $\rho_{11}=0.355\pm0.032,
    \rho_{22}=0.34\pm0.046,\rho_{33}=0.305\pm0.035$.
    \par
\begin{figure}[!ht]
\begin{center}
\begin{tabular}{c}
\includegraphics[
height=5cm ]{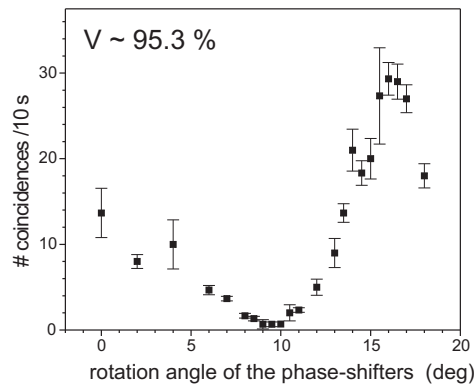}
\end{tabular}
\end{center}
\caption{\label{STangle} Space-time interference pattern for state
$|\Psi^\prime\rangle=\frac{1}{\sqrt{2}}(|2,0\rangle+e^{i\phi_{13}}|0,2\rangle)$.
Minimum corresponds to $\phi_{13}=\pi$, maximum corresponds to
$\phi_{13}=0$.}
\end{figure}

\begin{figure}[!ht]
\begin{center}
\begin{tabular}{c}
\includegraphics[
height=5cm ]{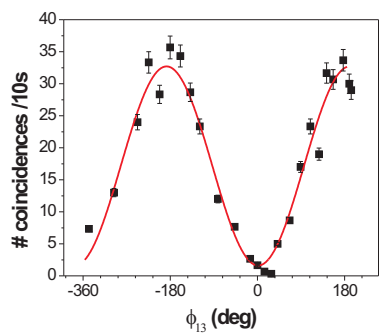}
\end{tabular}
\end{center}
\caption{\label{STphase} Space-time interference pattern for state
$|\Psi^\prime\rangle=\frac{1}{\sqrt{2}}(|2,0\rangle+e^{i\phi_{13}}|0,2\rangle)$.
Recalculated dependence on the phase $\phi_{12}$.}
\end{figure}

 {\bf Phase $\phi_{13}$.} The relative phase $\phi_{13}=\phi_{3}-\phi_{1}$ between states $|2,0\rangle$ and $|0,2\rangle$ can be controlled
    with the help of rotating quartz plates (QP2). The resulting state of biphoton field after these plates can be
    written as $|\Psi^\prime\rangle = |2,0\rangle+e^{i\phi_{13}}|0,2\rangle$. Varying the phase $\phi_{13}$ by rotating
    the quartz plates (QP), one can observe an interference pattern when the polarizers in Brown-Twiss scheme are set to
    transmit $+45^\circ$ and $-45^\circ$ polarized light \cite{olya:01} . This effect, known as "space-time interference",
    can be used to calibrate $\phi_{13}$, since we can assign value "0" to the position of the minimum of interference
    pattern, and "$\pi$" to the position of the maximum.  An obtained typical dependence from this
    parameter is presented at (Fig.~\ref{STangle}). A nonperiodic behavior is caused by a nonlinear dependence of an induced
    phase from the rotation angle. One can recalculate the dependence from the angle to dependence from the
    phase and the result is presented at (Fig.~\ref{STphase}).
    This plot serves as a calibration curve for phase $\phi_{13}$.

\begin{figure}[!ht]
\begin{center}
\begin{tabular}{c}
\includegraphics[
height=5cm ]{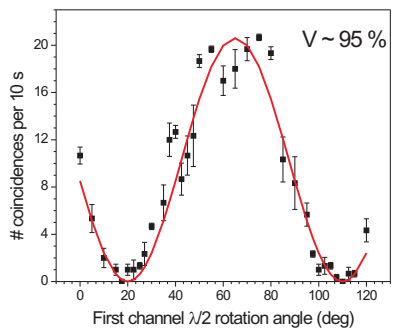}
\end{tabular}
\end{center}
\caption{\label{PolIntf} Polarization interference pattern from
compensated type II biphotons.}
\end{figure}

It is important to note that in order to ensure the resulting
state being close to a pure state it is necessary to
    achieve as high visibility as possible in space-time interference experiment with type I biphotons and polarization
    interference of type II biphotons (Fig.~\ref{PolIntf}). The obtained visibilities of approximately 95\%  were considered as a good result in
    order to proceed with a final stage of the experiment - the combination of states
    $|\Psi^\prime\rangle=|2,0\rangle+e^{i\phi_{13}}|0,2\rangle$ and $|1,1\rangle$ with a certain shift $\phi_{12}$
    between them.

    \par {\bf Phase $\phi_{12}$: pump interference.} The relation of the phase $\phi_{12}=\phi_{2}-\phi_{1}$ between
    the state $|\Psi^\prime\rangle$ and $|1,1\rangle$ to the voltage applied on PZT can be found by monitoring the pump
    interference pattern in M-Z interferometer. We found that the change of voltage by 1 V results in phase shift of
    $0.33 rad$ .

\begin{figure}[!ht]
\begin{center}
\begin{tabular}{c}
\includegraphics[
height=5cm ]{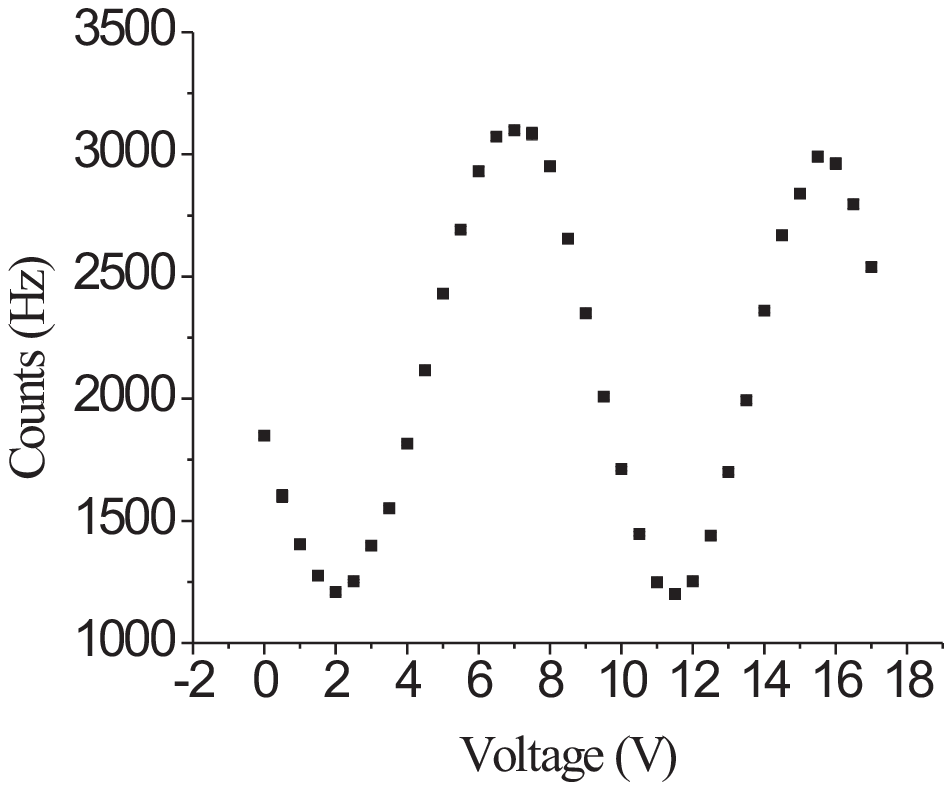}
\end{tabular}
\end{center}
\caption{\label{PumpInt} Pump interference pattern in Mach-Zehnder
interferometer. The distance between two maxima (in Volts)
corresponds to $\phi_{12}=\pi$.}
\end{figure}

   \par
\begin{figure}[!ht]
\begin{center}
\begin{tabular}{c}
\includegraphics[height=5cm]{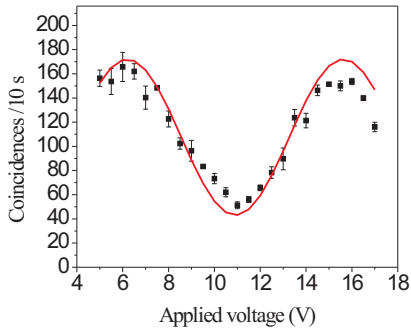}
\end{tabular}
\end{center}
\caption{\label{TypeI&TypeII} The interference pattern for state
$|\Psi^{out}\rangle$. The distance between two maxima (in Volts)
also corresponds to $\phi_{12}=\pi$.}
\end{figure}

 {\bf Phase $\phi_{12}$: generation of superposition between $|2,0\rangle$ and $|1,1\rangle$ states.} We also measured an interference pattern for a state
    $|\Psi^{\prime\prime}\rangle=\frac{1}{\sqrt{2}}(|2,0\rangle+exp(i\phi_{12})|1,1\rangle)$ varying the phase $\phi_{12}$.
    The polarization filters in Brown-Twiss scheme were set to measure
    moment $A$ of the tomography protocol. Moment $A$ exhibits no dependence on the phase $\phi_{12}$
    for a given state. Therefore we inserted an additional polarization transformer - a zero-order quarter waveplate
    operated at 800 nm and oriented at $20^{\circ}$ with respect to the vertical direction. The unitary action of this waveplate
    is described by $3x3$ matrix that converts an input state $|\Psi^{\prime\prime}\rangle$ to another state that
    is sent to a measurement scheme \cite{quarks} . For this certain waveplate, the state obtained after a transformer becomes
\begin{center}
\begin{equation}
\Psi^{out}=
\begin{pmatrix}
(-0.25+0.32i)exp(i\phi_{12})-0.15 \\
0.41iexp(i\phi_{12})+0.25+0.32i\\
(0.25+0.32i)exp(i\phi_{12})+0.15-0.54i
\end{pmatrix} \label{eq:rho}
\end{equation}
\end{center}
and now moment $A=c_{1}^{*}c_{1}$ shows a certain dependence on
$\phi_{12}$. The experimentally obtained dependence
(Fig.~\ref{TypeI&TypeII})
    can also be used to calibrate the phase. It is well known
    that a period of this interference pattern coincides with the
    the pump interference period. The theoretical visibility of the interference is equal to
    $64\%$, while the visibility obtained in the experiment is $53\%$. We attribute
    this difference to the misalignments of the setup.  \par We want to point out that in
    this experiment (for calibration purposes) we prepared a very specific
    superposition of basic states, interfering SPDC from type I
    and type II crystals, that was never reported in literature up
    to our knowledge.

   \subsection{Experimental procedure}
    In order to create a given qutrit state we needed to have independent control over four real parameters -
    two relative amplitudes and two relative phases. In the experiment we used HWP1 to control the amplitude of the
    state $|1,1\rangle$, and HWP2 to control the relative amplitudes of the states $|2,0\rangle$ and $|0,2\rangle$.
    The relative phase $\phi_{13}=\phi_{3}-\phi_{1}$ between the states $|2,0\rangle$ and $|0,2\rangle$ can be controlled
    with the help of rotating quartz plates (QP2). The relation of the phase $\phi_{12}=\phi_{2}-\phi_{1}$ between
    the state $|\Psi^\prime\rangle=|2,0\rangle+e^{i\phi_{13}}|0,2\rangle$ and $|1,1\rangle$ to the voltage applied to PZT can be found by monitoring the pump
    interference pattern in M-Z interferometer. We found that the change of voltage by 1 V resulted in the phase shift of
    $51.7^\circ$ and $\phi_{12}$ grew linearly with the applied voltage.
    \par States that constitute the first basis are trivial (Table I). They can be produced with the help of a single
    crystal, corresponding to type I or type II interaction. State $|2,0\rangle$ is generated when first $\lambda/2$ (HWP1) angle corresponds to the maximal reflection
    of the pump beam into the lower arm of a Mach-Zehnder and the angle of the second half-lambda waveplate (HWP2) is equal to $0^{\circ}$.
    In order to generate state $|0,2\rangle$, the HWP2 must be rotated by $45^{\circ}$ degrees from $0^{\circ}$,
    and to generate state $|1,1\rangle$ the HWP1 is rotated such, that the whole pump goes into the upper arm of Mach-Zehnder.
    Therefore, in the following, we will consider only the generation of the rest nine states, i.e. those forming the other three bases.
    According to Table I, only the relative phases
    between the basic states are to be varied. This allows us to use the same settings of the HWP's for the
    generation of nine states. It is also convenient to perform three sets of data acquisition - for the fixed
    $\phi_{13}$ values of $0$, $+120^\circ$ and $-120^\circ$, we change $\phi_{12}$ values in the range of, say, few
    periods and perform all tomographic measurements for each value of the phase $\phi_{12}$. Then we select the values
    of $\phi_{12}$ that correspond to the generation of the required state. For example, in order to generate the state
    $\beta^\prime$, we use $\phi_{13}=-120^\circ$ and $\phi_{12}=120^\circ$. The values of the moments at this point
    allow us to restore a raw density matrix of the generated state and compare it to the theoretical value.

    \par The following procedure was used in order to verify the orthogonality of the states that form a certain basis.
    First we chose a set state to which we would tune our polarization filters. Then the values of the angles of
    quarter- and half- waveplates (Fig. ~\ref{setup2}) ($\chi_1,\theta_1,\chi_2,\theta_2$) that assure the maximal projection of the polarization state
    of each photon on the $V$ direction can be calculated by mapping the set state on the Poincare
    sphere. Here, the lower index "1" corresponds to the transmitted
    arm, and the index "2" to the reflected arm of BS
    We chose states $|\alpha^\prime\rangle$, $|\alpha^{\prime\prime}\rangle$ and $|\alpha^{\prime\prime\prime}\rangle$
    to be our set states for each basis. Then, by setting the phase $\phi_{13}$ fixed and by varying the
    phase $\phi_{12}$ we measured the number of coincidence counts that correspond to the certain fourth order moment of
    the field. According to the orthogonality criterion, the coincidence rate should fall to zero when the values
    of $\phi_{13}$ and $\phi_{12}$ correspond to the generation of the states orthogonal to the set ones.

  \subsection{Results  and discussion} 
\begin{figure}[!ht]
\begin{center}
\includegraphics[height=8cm]{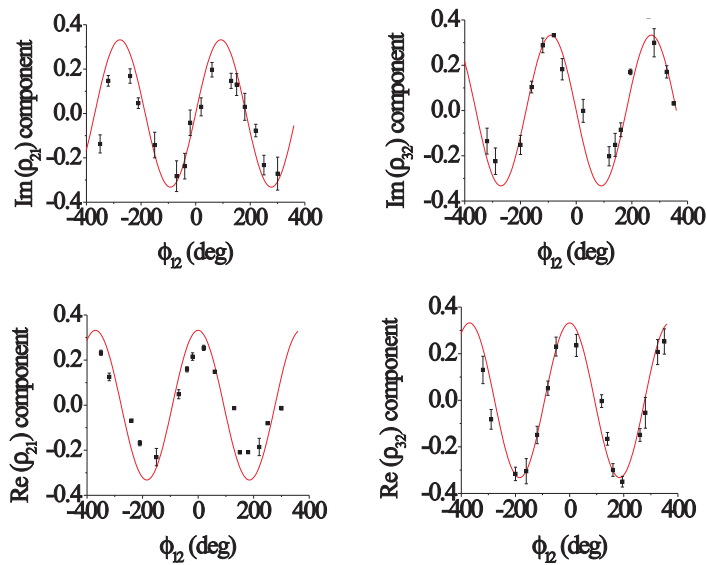}
\caption{Imaginary and real values of non-diagonal density matrix
components used to reconstruct state
$|\beta^{\prime\prime}\rangle$. Theoretical dependence is plotted
with a solid curve.} \label{Res}
\end{center}
\end{figure}

Let us consider the generation of the state $|\beta^{\prime\prime}\rangle$. In this case
    $\phi_{13}=0, \phi_{12}=120$. In Fig. ~\ref{Res} the measured values of the real and imaginary parts of the density matrix
    components $\rho_{21}$ and $\rho_{32}$ on phase $\phi_{12}$ are shown as function of the phase
    $\phi_{12}$. The number of accidental coincidences was negligibly
    small and was not subtracted in data processing.

    The phase $\phi_{13}=0$ remained constant during the tomography procedure. After obtaining
    the dependence of the moments $\rho_{21}$ and $\rho_{32}$ on phase $\phi_{12}$ we fitted our data with
    theoretical dependencies, using the least-square approximation method. The obtained values of
    all components were substituted in Eq.~\ref{eq:rho}. The obtained density matrix for state $|\beta^{\prime\prime}\rangle$
    is given below.

\begin{center}\footnotesize
\begin{equation}
\rho_{\beta^{\prime\prime}} = \left( \begin{array}{ccc}
0.355&-0.054-0.210i&0.315-0.010i\\
-0.054+0.210i&0.340&-0.106+0.262i\\
0.315+0.010i&-0.106-0.262i&0.305
\end{array}  \right)
\label{eq:beta}
\end{equation}
\end{center}

    The eigenvalues of this matrix are $\lambda_{1}=0.877, \lambda_{2}=0.136, \lambda_{3}=-0.013$.
    A corresponding set of eigenvectors is $X = (0.587,-0.173+0.521i,0.594-0.071i); Y =(0.642,0.379-0.649i,0.048+0.143i);
    Z =(0.493,-0.287+0.224i,-0.769-0.178i)$.
    Although the density matrix (Eq.~\ref{eq:beta}) is Hermitian and the condition $Tr(\rho)=1$ is satisfied, it doesn't correspond to any
    physical state because of the negativity of one of the eigenvalues. We want to point out
    that a first main component $(\rho_{exp}^{1})_{ij}=X_{i}X_{j}^{\ast}$ of a considered density matrix, which has a
    weight $0.878$ is already close to the theoretical state vector
    $|\beta^{\prime\prime}\rangle$ and the corresponding fidelity
    is $F=Tr(\rho_{th}\rho_{exp}^{1})=0.9903$.  The other two components correspond to the
    "experimental noise" that is due mainly to misalignments of a
    setup and small volume of collected data. We have obtained similar eigenvalues for all other states
    and raw fidelity computed for the main density matrix component as described above have varied
    from 0.983 to 0.998. We also employed the maximum likelihood method of quantum state root
    estimation (MLE) \cite{bogdan:03a,bogdan:03b} to make a tomographically reconstructed
    matrix satisfy its physical properties, such as positivity.
    The results are presented in the following table (Table II). The
    level of statistical fluctuations in fidelity estimation was
    determined by the finite size of registered events $(\sim 500)$.
    All experimental fidelity values lie within the theoretical range
    of $5\% (F=0.9842)$ and $95\% (F=0.9991)$ quantiles
    \cite{quant,bogdan:03a,quant} .

\begin{table}[!ht]
\begin{center} \footnotesize
\begin{tabular}{|c|c|c||c|c|c|}\hline
State&$F_{MLE}$&State&$F_{MLE}$&State&$F_{MLE}$\\\hline
$|\alpha^\prime\rangle$&$0.9989$&$|\alpha^{\prime\prime}\rangle$&$0.9967$&$|\alpha^{\prime\prime\prime}\rangle$&$0.9883$\\
\hline
$|\beta^\prime\rangle$&$0.9967$&$|\beta^{\prime\prime}\rangle$&$0.9989$&$|\beta^{\prime\prime\prime}\rangle$&$0.9989$\\
\hline
$|\gamma^\prime\rangle$&$0.9883$&$|\gamma^{\prime\prime}\rangle$&$0.9883$&$|\gamma^{\prime\prime\prime}\rangle$&$0.9967$\\
\hline
\end{tabular}
\caption{Fidelities estimated with Maximum Likelihood Method.}
\end{center}
\end{table}

\begin{figure}[!ht]
\begin{center}
\includegraphics[height=5cm ]{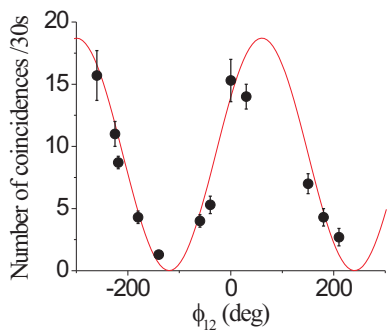} \caption{Dependence of number of
coincidences on a phase $\phi_{12}$ for a given settings of
polarization filters.} \label{ort}
\end{center}
\end{figure}

The obtained fidelity values show the high quality of the prepared
states.  The other test of the quality of states is the
fulfillment of the orthogonality criterion for the states that
belong to the same basis. For each set state we calculated the
settings of waveplates in our measurement setup that ensured the
maximal projection of each photon on the vertical polarization
direction. In Fig.~\ref{ort} we show the dependence of the
coincidence rate for the following setting of waveplates
$\chi_1=28.3^\circ,\theta_1=-33.5^\circ,\chi_2=-24^\circ,\theta_2=-2^\circ$.

These values correspond to the set state
$|\alpha^{\prime\prime\prime}\rangle$. As one can see, for the
fixed value $\phi_{13}=0$ the coincidence rate is almost equal to
zero, when phase $\phi_{12}=-120^\circ$. This corresponds to the
generation of the state $|\beta^{\prime\prime\prime}\rangle$,
which is orthogonal to $|\alpha^{\prime\prime\prime}\rangle$. The
visibility of this pattern is equal to 93.2\%. For the other
bases, the obtained values of visibilities varied from 92\% to
95\%. With these values of visibility, the lowest value of
coincidence rate corresponds to the accidental (Poissonian)
coincidence level and therefore the obtained data verifies the
orthogonality criterion.
\par Here we try to clarify how we understand the role of statistics in quantum state reconstruction.
When we find that fidelity F = 0.990, then this result is
sufficient without pointing out +/-0.XXX, because the obtained
value shows only the degree of correspondence between the
desirable result (to obtain fidelity as close to unity as
possible) and the achieved precision of quantum state
reconstruction. As usual, all works on quantum state tomography
end up on this. At the same time we think that if we consider the
question more thoroughly, it is necessary not only to point out
the result, but also to try to answer the question about the
principal precision that we can obtain in the experiment if we
consider the finite (and not so large) volume of statistical data.
Luckily we can also answer this question; it is overviewed in the
paper \cite{bogdan:03a} . On the basis of theory of statistical
fluctuations of state vector estimation, that was developed in
\cite{bogdan:03a} , we obtained that the interval of the expected
statistical fidelity fluctuations lies within 0.9842  (5\%
quantile) and 0.9991  (95\% quantile).
\par The fact that all obtained fidelity values lie within that
interval shows the correspondence of the experiment to the
statistical theory. We also point out that due to the complexity
of experiment, the volume of the data is not large (about 500
events in each of nine experiments). If we increase the volume of
data, then, sooner or later, it will come to the point, when the
correspondence of experiment to statistical theory will be
violated. The experimentally achieved fidelity value would be
lower then the theoretical due to the inevitable instrumental
errors of experiment (see also \cite{bogdan:03a}) . In this case
the quality of experiment can be estimated by the so called
"coherence volume" of data. If we exceed that volume, then the
fidelity "saturates" on the level which is somewhat less then
unity. All further increase of the volume will not lead to the
increase of fidelity since the precision of state reconstruction
will be limited by the instrumental errors and instability of the
experimental setup.
\section{Conclusions}
 We realized an interferometric method of
preparing the three-level quantum optical systems, that relied on
the polarization properties of single-mode two-photon light. The
specific sequence of states was generated and measured with high
fidelity values. The orthogonality of the states that form
mutually unbiased bases was experimentally verified. As an
advantage of this method we note that all control of the
amplitudes and phases of each basic state in superposition
(\ref{eq:state}) is done using linear optical elements, making it
easy to switch from one state to another and providing the full
control over the state (\ref{eq:state}) . The main disadvantage is
that we cannot generate an entangled qutrits in this
configuration. Our setup also allows one to prepare an arbitrary
polarization qutrit state on demand.

\acknowledgments Useful discussions with A.Ekert, B.Englert,
D.Kazlikowski, C.Kurtsiefer, L.C.Kwek, A.Lamas-Linares and A.Penin
are gratefully acknowledged. This work was supported in part by
Russian Foundation of Basic Research (projects 03-02-16444 and
02-02-16843) and the National University of Singapore's Eastern
Europe Research Scientist and Student Programme.

\end{document}